\documentclass[twocolumn,showpacs,amsmath,amssymb]{revtex4}
\usepackage{CJK}
\usepackage{amsmath}
\usepackage{graphicx}
\usepackage{amsfonts}
\usepackage{dcolumn}
\usepackage{bm}

\newcommand \dCG{$d$-CG\ }

\begin{document}

\title{Statistically consistent coarse-grained simulations
for critical phenomena in complex networks}

\author{Hanshuang Chen}

\author{Zhonghuai Hou}\email{hzhlj@ustc.edu.cn}

\author{Houwen Xin}

\author{YiJing Yan}

\affiliation{Hefei National Laboratory for Physical Sciences at
 Microscales \& Department of Chemical Physics, University of
 Science and Technology of China, Hefei, Anhui 230026, China}

\date{\today}

\begin{abstract}
We propose a degree--based coarse graining approach that not just
accelerates the evaluation of dynamics on complex networks, but also
satisfies the consistency conditions for both equilibrium
statistical distributions and nonequilibrium dynamical flows. For
the Ising model and Susceptible-Infected-Susceptible epidemic model,
we introduce these required conditions explicitly and further prove
that they are satisfied by our coarse-grained network construction
within the annealed network approximation. Finally, we numerically
show that the phase transitions and fluctuations on the
coarse-grained network are all in good agreements with those on the
original one.
\end{abstract}
\maketitle

\section{Introduction} \label{sec1}
Complex network has been one of the most active research topics in
statistical physics and many other disciplines
\cite{RMP02000047,AIP02001079,SIR03000167,PRP06000175,PRP08000093}.
It describes not only the pattern discovered ubiquitously in real
world but also a unified theoretical framework to understand the
inherent complexity in nature. However, the investigation of large
networks, such as human brain that composes of about $10^{11}$
neurons and $10^{14}$ synapses \cite{RMP06001213}, requires
tremendous time-demanding efforts. The phenomenological description
may capture certain properties of system, but always neglects
microscopic information. A promising alternative is to develop
coarse-grained (CG) methods, aiming at significant reducing the
degree of freedom while proper preserving the microscopic
information of interest.

Several CG schemes have been proposed. Renormalization
transformation has been used to simplify self--similar networks, and
the reduced networks often preserve some topological properties of
the original ones
\cite{PRL04016701,NTR05000392,PRL06018701,PRL08148701}. Spectral
coarse graining technique has been proposed, in which the
eigenvalues of Laplace matrix of network are almost unchanged, such
that the dynamics of random walk and synchronization are preserved
\cite{PRL07038701,PRL08174104}. Equation--free multiscale
computational framework has been developed to accelerate simulation
using a coarse time stepper \cite{CMS03000715}. This approach has
been applied to study the CG dynamics of oscillators network
\cite{PRL06144101}, gene regulatory network \cite{JCP06084106}, and
adaptive epidemic network \cite{EPL08038004}. 
However, to our best knowledge, no attempt has been made for
developing CG simulation method to study critical phenomena (usually
described by stochastic models) on complex networks. 
The size--dependent and scaling behaviors in these systems are
studied so far mainly by such as Monte Carlo (MC) and kinetic MC
(KMC) simulations \cite{RMP08001275,AIP00000815}. Apparently, these
microscopic approaches are often too expensive. It is noticed that
the CG stochastic models have been proposed to study
reaction--diffusion processes on regular lattices
\cite{JCP03000250}. However, the existing methods are largely
inapplicable to critical phenomena on complex networks with
diversified heterogeneity. Moreover, the crucial issue concerning
the methodology development as to what criteria should be met to
make the CG model statistically consistent with the microscopic one
is yet to be addressed.

In this paper, we propose the degree-based CG (\dCG for short
hereafter) be a statistically consistent scheme, within the annealed
network approximation (ANA)
\cite{RMP08001275,PRE03036112,PRL02258702}. It may therefore be an
efficient and reliable CG method for evaluating the stationary
properties of phase transitions and studying size effect on complex
networks.
We put forward the conditions of statistical consistency (CSC) on
both equilibrium and nonequilibrium properties, exemplified  with
the Ising model and Susceptible--Infected--Susceptible (SIS) model,
respectively. We show that the \dCG approach that merges together
the nodes of similar degrees does warrant the CSC within ANA. The
stochastic Ising spin--flip and epidemic spreading dynamics can
therefore be evaluated faithfully and efficiently with the \dCG
networks. The calculated phase diagrams, fluctuation dynamics, and
system size--scaling behaviors are all shown in excellent agreements
with the corresponding microscopic MC and KMC results.

The rest of this paper is organized as follows. In Sec.\ref{sec2},
we present a general scheme for coarse graining network and \dCG
approach, and further prove that this approach satisfies statistical
consistency. Extensive numerical demonstrations of the CG approach
are performed on diverse networks in Sec.\ref{sec3}. At last, main
conclusions and discussion are addressed in Sec.\ref{sec4}.

\section{Coarse Graining Procedure}  \label{sec2}

\subsection{Network Coarse Graining}  \label{sec2.1}
\begin{figure}
\centerline{\includegraphics*[width=0.8\columnwidth]{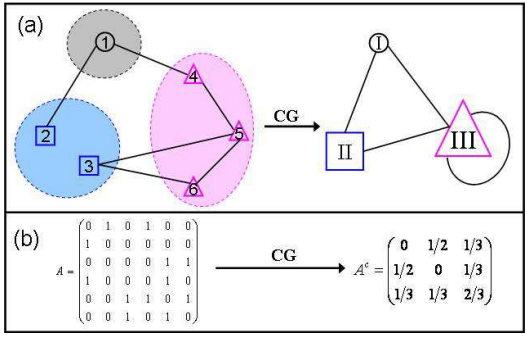}}
\caption{(color online) (a) A schematic example of coarse graining
network in which nodes of the same shape are merged; (b) the
adjacency matrix ${\bm A}^c$ of CG network by Eq.\,(\ref{eq1}). Note
that the CG network may include self--connections; for example, the
CG-node III where the connected nodes are merged. \label{fig1}}
\end{figure}

Let us start with the basic ingredients of network coarse graining.
Consider a network consisted of $N$ nodes whose connectivity is
given by the adjacency matrix ${\bm A}$, in which $A_{ij}= 1$ if
nodes $i$ and $j$ are connected and $A_{ij}=0$ otherwise. Merging
$q_\mu$ nodes together into a CG-node (denoted by $C_\mu$) leads to
a CG--network with $N^c$ CG-nodes. 
We adopt the mean--field definition of the CG connectivity between $C_\mu$ and $C_\nu$, as the average
number of links connecting any two nodes inside $C_\mu$ and $C_\nu$. The adjacency matrix ${\bm A}^c$ of the
CG network is then
\begin{equation}
 A_{\mu\nu}^c  =
  \begin{cases}
   \frac{2}{q_\mu(q_\mu-1)}\sum\limits_{i,j \in C_\mu; i<j} A_{ij}
  & \text{if $\mu = \nu$},  \\
  \frac{1}{q_\mu q_\nu}\sum\limits_{i \in C_\mu,j \in C_\nu} A_{ij}
  & \text{if $\mu \ne \nu$}.
 \end{cases}
\label{eq1}
\end{equation}
For illustration, Fig.\ref{fig1} depicts an example of coarse
graining a network with six nodes into a weighted CG network with
three CG-nodes.

\subsection{CSC} \label{sec2.2}
We address the issue on the statistical consistency of a CG scheme
with the microscopic network, in terms of both the equilibrium
distribution and the nonequilibrium flow. We exploit the Ising model
and the SIS model as the paradigmatic examples for equilibrium and
nonequilibrium systems, respectively.

For the Ising model defined on a network, the Hamiltonian is given
by $H=-J\sum\nolimits_{i < j}{A_{ij}s_i s_j}$, with the spin
variable $s_i=\pm 1$, and the ferromagnetic interaction parameter
$J>0$. The probability of a given microscopic spin configuration
$\{s_i\}$ is given by canonical distribution $e^{-\beta H}/Z$, where
$\beta$ is the inverse temperature and $Z$ is the partition
function. The corresponding CG Hamiltonian assumes the sum of
pairwise interactions inside and between the CG-nodes,
i.e.~$H^c=H_1^c+H_2^c$,
\begin{align}
 H_1^c
&= -\frac{J}{2}\sum\limits_\mu {A_{\mu\mu}^c}
   \big[n^+_\mu(n^+_\mu - 1) + n^-_\mu(n^-_\mu - 1) - 2n^+_\mu n^-_\mu\big]
\nonumber \\
&= -\frac{J}{2}\sum\limits_\mu {A_{\mu\mu}^c}
   \big(\eta^2_\mu - q_\mu \big),
\label{eq2} \\
 H_2^c
&= -J\sum\limits_{\mu<\nu} A^c_{\mu\nu}
   \big(n^+_\mu n^+_\nu + n^-_\mu n^-_\nu - n^+_\mu n^-_\nu - n^-_\mu n^+_\nu\big)
\nonumber \\
&= - J\sum\limits_{\mu<\nu} A^c_{\mu\nu} \eta_\mu \eta_\nu . \label{eq3}
\end{align}
Here, $\eta_\mu=\sum\nolimits_{i \in C_\mu }{s_i}$ is the CG spin variable, and $n_\mu^\pm={{\left( {q_\mu
\pm \eta_\mu }\right)}\mathord{\left/ {\vphantom{{\left( {q_\mu \pm \eta_\mu}\right)}2}}\right.
\kern-\nulldelimiterspace}2}$ is the number of up/down spins inside $C_\mu$. Note that $H^c=H^c_1+H^c_2$
above is a closure expression at the CG level that depends only on ${\bm A}^c$ and $\{\eta_\mu\}$. 
 To make the CG model consistent with
the microscopic one, it demands that the probability of any given CG configuration $\{\eta_\mu\}$ in
equilibrium be the sum of the probabilities of all microscopic configurations that contribute to it. That is
\begin{equation}
  g(\{\eta _\mu\}) e^{-\beta H^c}
= \sum_{\{s_i\}} \prod_\mu
  \delta\Big(\eta_\mu  - \sum_{i \in C_\mu }\!\!s_i\Big) e^{-\beta H} ,
\label{eq4}
\end{equation}
where $g(\{\eta_\mu\})= \prod_\mu {q_\mu}! \big/ ({n^+_\mu}! {n^-_\mu}!)$ is the number of microscopic
configurations corresponding to $\{\eta_\mu\}$.

We now consider the nonequilibrium scenario, exemplified by the SIS
network for epidemic spreading dynamics. 
At the microscopic level, the SIS network nodes represent
individuals being either susceptible ($\sigma=0$) or infected
($\sigma=1$).
A susceptible individual $j$ can get infected at rate $\tilde r_j=\sum\nolimits_i f_{i \to j}$, where
$f_{i\to j}=\lambda A_{ij}\sigma_i(1-\sigma_j)$ is the spreading flow from an infected individual $i$ to
$j$. On the other hand, an infected node can recover to susceptible state at rate $r_i=\gamma\sigma_i$.
Without loss of generality, we set $\gamma=1$  hereafter to scale the infection parameter $\lambda$. The SIS
model exhibits a nonequilibrium dynamical phase transition at $\lambda=\lambda_c$ from an absorbing state
(all recovered) to an active state (disease spreading persistently) \cite{AIP00000815,JRS05000295}. For the
CG--SIS network, we define the CG variables as the number of infected individuals inside a CG-node $C_\mu$
by $\sigma^c_\mu= \sum_{i \in C_\mu}{\sigma_i}$. 
Written in a closure form, the CG recovery rate is $r^c_\mu =\sigma^c_\mu$, and the CG spreading flow
between two CG-nodes is $f^c_{\mu \to \nu}=\lambda A^c_{\mu\nu}\sigma^c_\mu ({q_\nu-\sigma^c_\nu})$. For
this nonequilibrium system, the CG model is consistent with the microscopic one if the CG flow matches the
microscopic flows:
\begin{equation}
  f_{\mu \to \nu}^c  = \sum_{i \in C_\mu, j \in C_\nu} f_{i \to j} \label{eq5}
\end{equation}
Note that the CG recovery rate of $r^c_\mu = \sum_{i \in C_\mu} {r_i}$ holds trivially.

\subsection{Degree-based CG Scheme} \label{sec2.3}
In our approach, we merge the nodes with similar degrees together.
We shall show the resulting \dCG scheme does satisfy the CSC, as
defined by Eq.\,(\ref{eq4}) for the Ising model and Eq.\,(\ref{eq5})
for the SIS model, within the ANA for the ensemble averaged dynamics
\cite{RMP08001275,PRE03036112,PRL02258702}. In many previous studies
\cite{PLA02000166,PRE02035108}, ANA has been extensively confirmed
to be a useful tool for describing quenched networks, as in case of
the present paper. Although ANA is just an approximation, it still
gives a reasonable description of the average behavior of nodes of
the same degree. Moreover, one can consider that the ANA is a
statistical characterization of large number of quenched networks
with the same degree distributions. Nevertheless, in
Ref.\cite{PRE09051127} it has been pointed out that there exists
some discrepancies between considering annealed approximation for
quenched networks and considering annealed network models by
themselves. According to the ANA, one can replace the dynamics on a
given network of $N$ nodes by that on a weighted graph of the full
connectivity $A_{ij}=k_i k_j/(N{\langle k \rangle})$, where $k_i$
and $k_j$ are the degrees of nodes $i$ and $j$, respectively, and
$\left\langle k \right\rangle$ is the mean degree. In an ideal \dCG
scheme, the microscopic nodes in a single CG-node are of the same
degree: $k_i|_{i \in C_\mu}= K_\mu$. 
The CG connectivity is then $A^c_{\mu\nu}=\frac{K_\mu K_\nu}{\langle
k \rangle N}$, for the CG--network dynamics treated at the ANA
level. As results, Eqs.(\ref{eq2}) and (\ref{eq3}) become,
respectively (noting that $s^2_j = 1$)
\begin{align}
 H_1^c &= -\frac{J}{2N \langle k\rangle}
  \sum_\mu K_\mu^2 \Big[\sum_{i,j \in C_\mu} s_i s_j -q_\mu \Big]
\nonumber \\
&=  -\frac{J}{N \langle k \rangle}
 \sum_\mu {K_\mu^2} \sum_{i,j \in C_\mu; i < j} s_i s_j ,
\label{eq6}\\
 H_2^c &= - \frac{J}{N \langle k \rangle}
   \sum_{\mu<\nu} K_\mu K_\nu
   \sum_{i\in C_\mu, j\in C_\nu} s_i s_j .
\label{eq7}
\end{align}
Their sum can be written as
\begin{equation}
 H^c = - \frac{J}{N\langle k \rangle} \sum_{\mu,\nu} K_\mu K_\nu
  \sum_{i\in C_\mu, j\in C_\nu; i < j} s_i s_j .
\label{eq8}
\end{equation}
It is identical to the microscopic Hamiltonian at the ANA level.
In other words, the Hamiltonian of any \dCG configuration equals to
the collection of its contributing microscopic configurations.
The pre-exponential terms in two sides of Eq.(\ref{eq4}) are both
the degeneracy of CG configuration, as well as the identical energy
factors ($H^c=H$).
 We have thus proved that
the ideal \dCG approach to the Ising model obeys the CSC of
Eq.(\ref{eq4}) exactly.
 It is also easy to prove that the nonequilibrium CSC of
Eq.(\ref{eq5}) is true for the SIS model in consideration; both
sides there equal to $\lambda\frac{K_\mu K_\nu}{N\langle k
\rangle}{\sigma^c_\mu}(q_\nu - \sigma^c_\nu)$. Therefore, the \dCG
approach satisfies the CSC for both equilibrium probability
distributions and nonequilibrium dynamical flows. Certainly it is
anticipated that the CSC holds approximately if merged together are
the nodes of similar degrees rather than the exactly same ones. 

\section{Numerical Demonstrations} \label{sec3}

\subsection{CG-MC and CG-KMC simulations} \label{sec3.1}
The MC simulation with the Metropolis dynamics \cite{Lan2000} and
the KMC simulation \cite{JCP77002340} are applied to the Ising model
and SIS model, respectively, at both the microscopic and the CG
levels. In the CG-MC simulation, a CG-node $C_\mu$ is randomly
chosen, followed by the Metropolis try for spin-flip process. The
probabilities of flipping a up/down--spin is $n^{\pm}_\mu W(\beta,
\Delta E_{\uparrow/\downarrow})$, where $W(\beta, \Delta
E)=\min(e^{-\beta \Delta E}, 1)$, and $\Delta
E_{\uparrow/\downarrow}$ is the change of CG Hamiltonian resulting
from the flip of a up/down spin. It is easy to confirm that the
CG-MC simulation obeys the detailed balance condition. In the CG-KMC
simulation, following the Gillespie algorithm \cite{JCP77002340}, a
CG process to be executed is randomly selected based on the
transition rates of all processes. The configuration and transition
rates are updated for executing the next CG process.

\begin{figure}
\centerline{\includegraphics*[width=0.8\columnwidth]{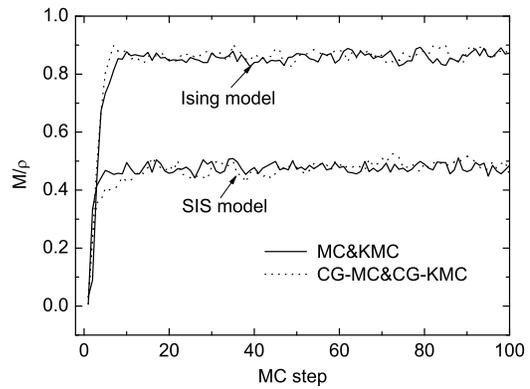}}
\caption{Typical time evolutions of the magnetization $m$ in Ising
model at $T=10$ (in unit of $J/k_B$) and the density of infected
nodes $\rho$ in SIS model at $\lambda=0.1$ for both microscopic and
CG levels. Other parameters are $N=1024$ and $N^c=16$. \label{fig2}}
\end{figure}

\subsection{Scale-free networks} \label{sec3.2}
We first consider the Barab\'{a}si--Albert (BA) scale-free network
\cite{SCI99000509}, with the degree distribution follows a power-law
$P(k)\sim k^{-\gamma}$ with scaling exponent $\gamma=3$. Figure
\ref{fig2} plots typical time evolutions of the magnetization
$m=\sum\nolimits_i {{s_i}}/N$ in Ising model at $T=10$ (in unit of
$J/k_B$) and the density of infected nodes $\rho=\sum\nolimits_i
{{\sigma_i}}/N$ in SIS model at $\lambda=0.1$, where $N=1024$ and
$N^c=16$ are used. For both the microscopic and CG simulations, the
systems attain the steady states associated with fluctuating noise
after transient time. It is clear that there are in good agreement
in the steady-state values of $m$ and $\rho$, as well as their
fluctuating amplitudes for both simulations cases.

\begin{figure}
\centerline{\includegraphics*[width=0.8\columnwidth]{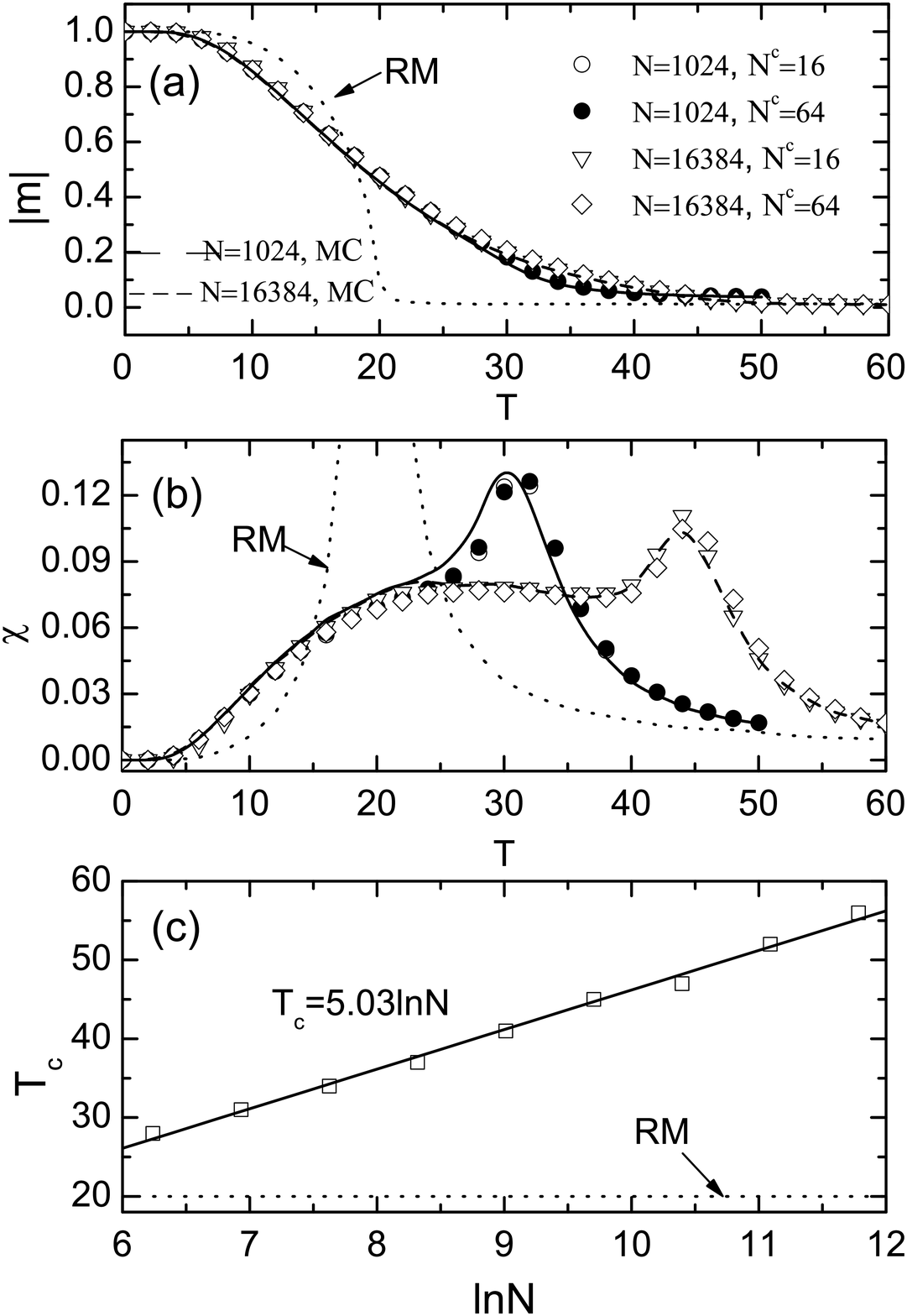}}
\caption{(a)-(b) $|m|$ and $\chi$ as a function of $T$. Symbols and
lines correspond to results by simulating the CG model and by direct
microscopic MC simulation, respectively. (c) $T_c$ as a function of
$\ln N$ obtained by the CG model with $N^c=64$. All dotted lines are
the results of randomly merging models with $N=16384$ and $N^c=64$.
All the networks have fixed mean degree $\langle k \rangle=20.$ The
error bars (not shown) are smaller than the symbol size.
\label{fig3}}
\end{figure}

\begin{figure}
\centerline{\includegraphics*[width=0.8\columnwidth]{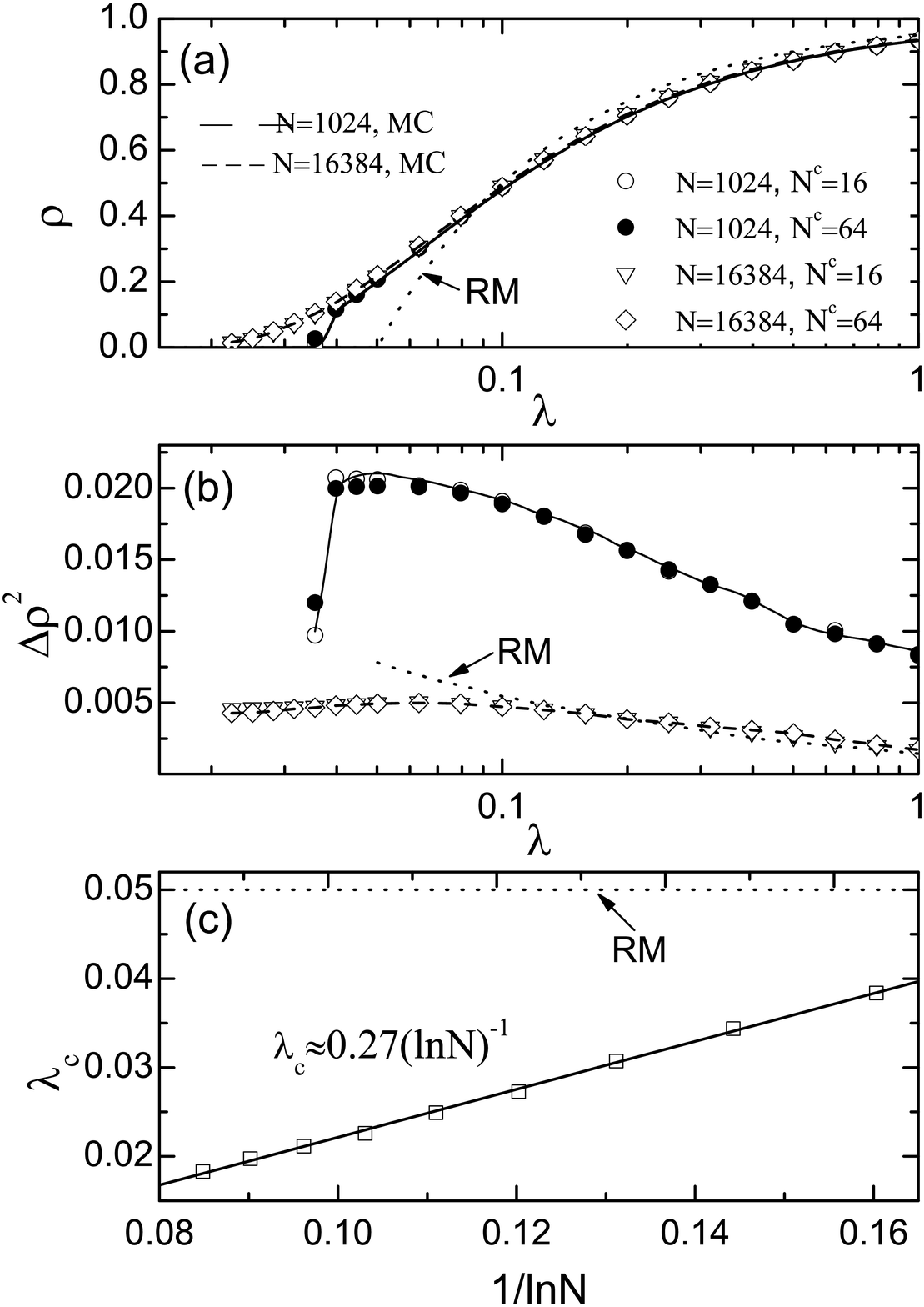}}
\caption{(a)-(b) $\rho$ and and $\Delta \rho^2$ as a function of
$\lambda$. Symbols and lines correspond to results by simulating the
CG model and by direct microscopic KMC simulation, respectively. (c)
$\lambda_c$ as a function of $1/\ln N$ obtained by the CG model with
$N^c=64$. All dotted line are the results of randomly merging models
with $N=16384$ and $N^c=64$. All networks have fixed mean degree
$\langle k \rangle =20$. The error bars (not shown) are smaller than
the symbol size. \label{fig4}}
\end{figure}

For the Ising model, we construct initial configurations by
preparing each node with a random spin value $s_i=+1$ or $-1$ with
an equal probability. As the simulation proceeds, the system quickly
relaxes to an equilibrium state. With the same temperature $T$ we
run at least $100$ times of simulations corresponding to different
initial configurations and network realizations. In each simulation,
$2\times10^3$ MC steps (MCS: each spin is attempted to flip once on
average during each MCS) are performed and the last $10^3$ MCS are
used to investigate the system's behavior. As $T$ decreases the
value of the magnetization $m$ undergoes a transition from zero to
nonzero at the critical temperature $T_c$. Below $T_c$ we notice
that due to finite-size effects the system can switch between two
stable states via a nucleation mechanism, resulting in the
oscillations of $m$ \cite{PHA02000260}. Above $T_c$, $m$ fluctuates
around zero, and the susceptibility $\chi$ per node has a maximum at
phase transition, which can be used to determine $T_c$ as we shall
show in Fig.\ref{fig3}(b). The susceptibility is related to the
magnetization fluctuation via the fluctuation--dissipation theorem.
To avoid the offset of $m$ due to its oscillations in simulations,
we will use instead the absolute value of $m$ in the following.
$|m|$ and $\chi$ are shown in Fig.\ref{fig3}(a) and (b),
respectively, as a function of the temperature $T$ for different $N$
and $N^c$. Apparently, the \dCG results (reported by symbols) are in
excellent agreements with the microscopic--level counterparts (by
solid and dashed lines). As contrast, we also report the results of
random-merging (RM) CG scheme (by dotted lines) in Fig.\ref{fig3}(a)
and (b), where each CG--node includes $q_\mu\equiv N/N^c$ ($N=1024$,
$N^c=64$) nodes selected randomly from the whole network. Evidently
the random scheme fails badly in reproducing the microscopic
behaviors. We also used the same $q_\mu$ in the random scheme as in
the case of \dCG scheme, and found that the two random schemes
produce the same results.

It is important to merge nodes with similar degrees together, as
already shown both analytically and numerically. Strikingly, even
when the original network is reduced to one with only $16$ CG-nodes,
the \dCG scheme still faithfully reproduces the phase transition
curves and fluctuations properties. Since $N^c$ is largely reduced
compared to $N$, a considerable speed--up of CPU time, about a
factor of 40, is realized for $N=16384$. A significantly higher gain
can be expected as the network gets larger, allowing the
computational study of network size effect very affordable.

Using the \dCG approach with a fixed size of CG--networks $N^c=64$,
we calculate the dependence of phase--transition critical $T_c$ on
the network size $N$, as reported in Fig.\ref{fig3}(c). It had been
well established that the Ising model with ferromagnetic
interactions on BA scale-free network undergoes a phase transition
from ferromagnetism to paramagnetism at a critical temperature $T_c$
that increases as the logarithm of network size
\cite{PHA02000260,PLA02000166,EPB02000191,PRE02016104}. Our result
is consistent with the theoretical expression \cite{PLA02000166}
that ${T_c}=\frac{\langle k \rangle}{4}\ln N$, where ${\left\langle
k \right\rangle }=20$ for the present case of study. Note that the
critical phenomenon disappears  when $N \to \infty$.

For the SIS model there is an epidemic threshold $\lambda _c$ on a
finite size BA scale-free network network
\cite{PRL01003200,PRE02035108}. Our numerical simulation starts from
a random configuration with about half nodes being infected. After
an initial transient regime, the system will evolve into a steady
state with a constant average density of infected nodes. The steady
density of infected nodes $\rho$ is computed by averaging over at
least $50$ different initial configurations and at least $10$
different network realizations with the same parameter $\lambda$.
The epidemic threshold $\lambda _c$ occurs at $\rho=0$ (absorbing
state) if $\lambda<\lambda _c$ and $\rho>0$ (active state) if
$\lambda>\lambda _c$ \cite{PRL01003200}. Due to finite size effects,
the fluctuation can drive the system to the absorbing state,
especially in the vicinity of $\lambda_c$. Once the absorbing state
is arrived, the system will never leave it. Based on the
consideration, we use, in practice, a nonzero tolerance in $\rho$
(with the order of $N^{ - 1}$) as the boundary of the phase
transition point. Reported in Fig.\ref{fig4}(a) and (b) are the
calculated results of $\rho$ and its fluctuation $\Delta {\rho ^2} =
\left\langle {{\rho ^2}} \right\rangle  - {\left\langle \rho
\right\rangle ^2}$, respectively, as a function of $\lambda$,
obtained by the CG model and the microscopic model. Again, while the
RM scheme fails when the relative infection rate parameter
$\lambda<0.1$, the agreement between the \dCG and the microscopic
results remains excellent. Fixing $N^c=64$ in the \dCG approach, the
resulting $\lambda_c$ shown in Fig.\ref{fig4}(c) is proportional to
$1/\ln N$, also consistent with the theoretical prediction
\cite{PRE02035108}.

\begin{figure}
\centerline{\includegraphics*[width=1.0\columnwidth]{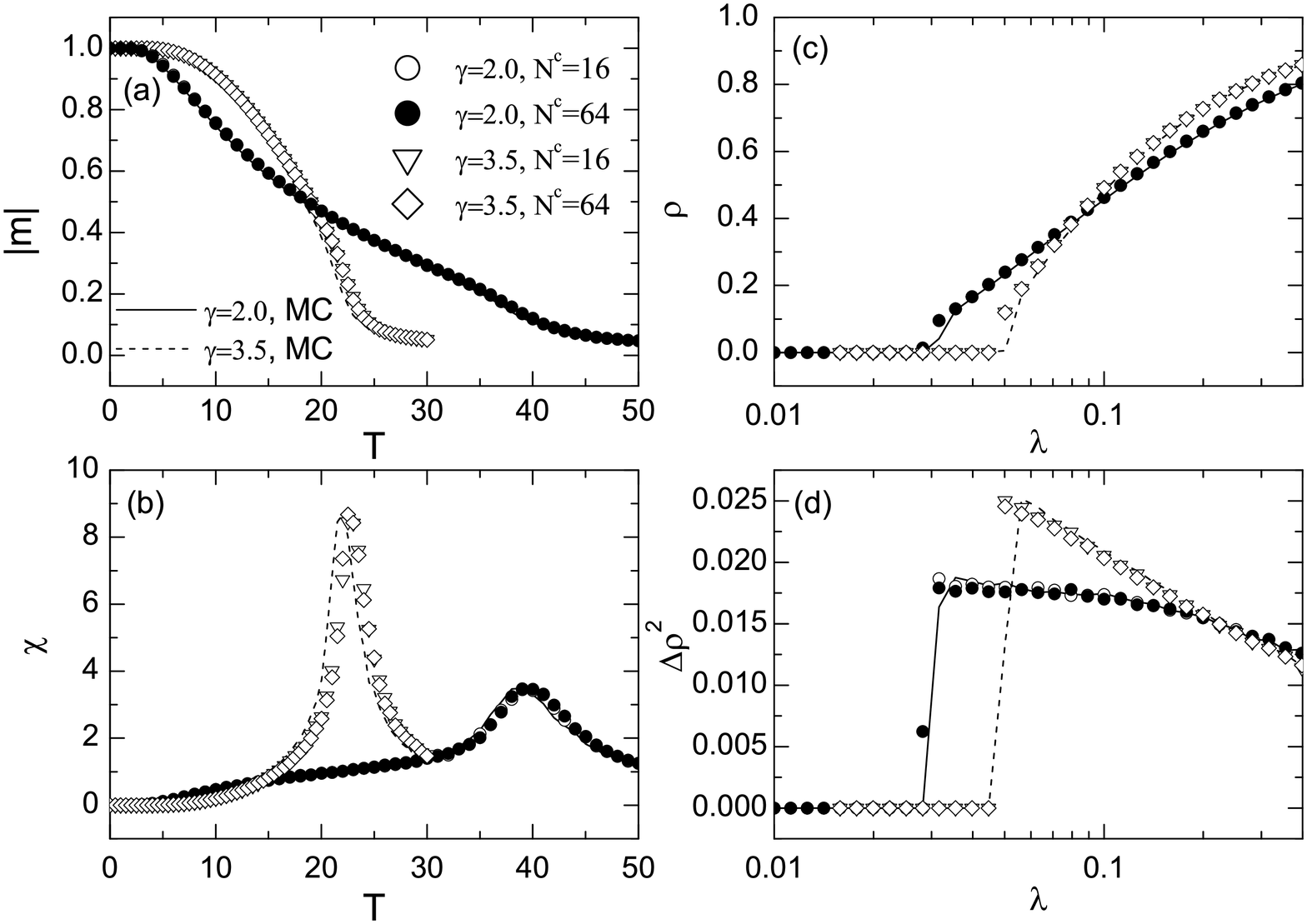}}
\caption{Comparison of microscopic and CG simulations on scale-free
networks with the scaling exponent $\gamma=2.0$ and $\gamma=3.5$.
Left panel: $|m|$ and $\chi$ as a function of $T$ for Ising model.
Right panel: $\rho$ and and $\Delta \rho^2$ as a function of
$\lambda$ for SIS model. All networks have fixed size of $N=1024$
and mean degree of $\langle k \rangle =20$. \label{fig5}}
\end{figure}

Figure \ref{fig5} show that the simulation results of the
microscopic and CG levels on scale-free networks with other two
scaling exponents, $\gamma=2$ and $\gamma=3.5$. For larger $\gamma$,
$T_c$ becomes lower, while $\lambda_c$ gets larger. It is clearly
observed that our \dCG approach is still applicable.  In addition,
many other types of networks such as random network and small-world
network are also tested, and all results show that the validity of
our \dCG approach does not depend on network topology.

\subsection{Degree correlated networks} \label{sec3.3}
\begin{figure}
\centerline{\includegraphics*[width=1.0\columnwidth]{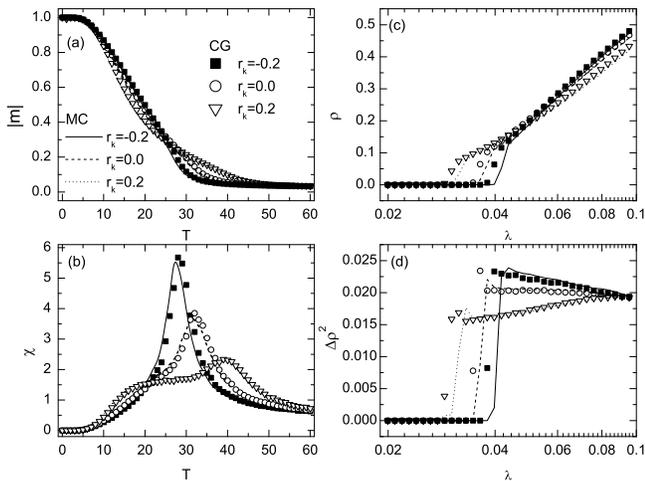}}
\caption{Comparison of microscopic and CG simulations on correlated
networks for different degree-mixing coefficient $r_k$. Left panel:
$|m|$ and $\chi$ as a function of $T$ for Ising model. Right panel:
$\rho$ and and $\Delta \rho^2$ as a function of $\lambda$ for SIS
model. All networks have power-law degree distributions with the
scaling exponent $\gamma=3$, and fixed size of $N=1024$ and mean
degree of $\langle k \rangle =20$. \label{fig6}}
\end{figure}

It is worthy noting that the above numerical demonstrations are
carried out on degree uncorrelated networks. We will show that our
\dCG approach is valid to reproduce critical behaviors on
degree--degree correlated networks as well. It has been witnessed
that many real networks display different degree--mixing patterns
\cite{PRL02208701}. To measure the degree of the correlation, in
Ref.\cite{PRL02208701} Newman introduced a degree-mixing
coefficient: $r_k  = ( {\left\langle {k_i k_j } \right\rangle -
\left\langle {k_i } \right\rangle \left\langle {k_j } \right\rangle
} )/( {\left\langle {k_i^2 } \right\rangle - \left\langle {k_i }
\right\rangle ^2 } )$, where $k_i$ and $k_j$ are the remaining
degrees at the two ends of a link and $\left\langle \bullet
\right\rangle$ means the average over all links. $r_k=0$ indicates
that there is no degree correlation, while $r_k>0$ ($<0$) indicates
that a network is assortatively (disassortatively) mixed by degree.
Previous studies have revealed that degree--mixing pattern plays an
important role in dynamical behaviors on networks, such as
percolation \cite{PRL02208701}, epidemic spreading
\cite{LNP03000127}, synchronization \cite{PRE06066107}. To generate
different degree--mixing networks, we employ a algorithm proposed in
\cite{PRE04066102}. At each elementary step, two links in a given
network with four different nodes are randomly selected. To get an
assortative network, the links are rewired in such a way that one
link connects the two nodes with the smaller degrees and the other
connects the two nodes with the larger degrees. Multiple connections
are forbidden in this process. Repeat this operation until an
assortative network is generated without changing the node degrees
of the original network. Similarly, a disassortative network can be
produced with the rewiring operation in the mirror method. We start
from BA scale-free networks with a neutrally degree-mixing pattern,
and produce some groups of degree-mixing networks by performing the
above algorithm. Figure \ref{fig6} displays the results of Ising
model and SIS model for three different values of $r_k$. For each
$r_k$, the simulation results of CG models agree well with those of
microscopic ones. In Ising model, $T_c$ shifts to right and $\chi$
at $T_c$ becomes smaller as $r_k$ increases. In SIS model, both
$\lambda_c$ and the fluctuation of $\rho$ at $\lambda_c$ decrease
with $r_k$.

\section{Conclusions and Discussion} \label{sec4}
In summary, we propose an approach for coarse graining the phase
transition dynamics on complex networks described by stochastic
models for both equilibrium and nonequilibrium systems. The \dCG
approaches via degree-based merging scheme are feasible since the
reliable microscopic information such as phase transition behaviors
and fluctuations are preserved. We have verified that our \dCG
approach supports, in consistent with the microscopic models, the
equilibrium distributions for Ising model and the nonequilibrium
dynamical flows for SIS model. Stochastic description, as
exemplified here, is ubiquitously important in the study of phase
transition dynamics and complex networks for a wide range of
realistic systems. Moreover, this work also suggests the development
of other promising CG statistical models satisfying CSC.

It is interesting to compare our \dCG approach with heterogeneous
mean-field theory (HMFT) that successfully predicts $T_c$ and
$\lambda_c$ on heterogeneous scale-free networks
\cite{PRE02016104,PRL01003200}. Based on the ansatz that nodes with
the same degree share the same dynamical properties, HMFT derives a
series of coupled mean-field equations for degree-dependent
quantities. Recently, Langevin approach together with the HMFT has
been developed in Refs.\cite{PRE09036110,JSM0910004}, which have
confirmed that such an approach is responsible for the effect of
fluctuations in a finite-size network that often play important
roles in the vicinity of phase transitions. With regard to the
present study, we develop a reliable CG simulation approach, that
not only can correctly predict the critical phenomena and
fluctuations information, but also can be applied to diverse
networks. Especially, it is shown that the validity of our \dCG
approach on the application of correlated networks; however, in this
case the HMFT becomes, in general, very difficult to deal with. This
is because that specific formulation of degree--degree correlation
are unknown for most of correlated networks, such as  $P(k'|k)$,
that is the conditional probability of a node of degree $k$ being
connected to a node of degree $k'$. On the other hand, the CSC
discussed in this work may provide a solid understanding of the
physical mechanism behind the basic assumption of HMFT. Our analysis
here may lead to the advancement in efficient and consistent CG
approaches for dynamics on surfaces and soft lattices.

Note that the present study is limited to the case of quenched
networks, that is, the connectivity of networks is frozen in time.
While for the case of annealed networks, i.e. the networks
themselves are dynamical objects, our CG approach will encounter
some difficulties in application. In this case, since the network
connections are frequently reshuffled, the adjacency matrix of the
resulting CG--network and CG variables should be accordingly
updated. This will lead to the very inefficiency of our CG approach
in simulations. However, an important advancement in studying
critical phenomena on annealed networks has been made in
Refs.\cite{PRE09056115, PRE09051127} by means of finite-size scaling
theory. Developing coarse-grained simulation methods on annealed
networks and adaptive networks \cite{JRS08000259} deserve further
investigations.

\begin{acknowledgments}
This work was supported by the National Science Foundation of China
under Grants No. 20933006 and No. 20873130.
\end{acknowledgments}

%

\end{document}